\documentclass[reprint,amsmath,amssymb,aps,showpacs,prb,superscriptaddress]{revtex4-1}
\usepackage[dvips]{graphicx}
\usepackage{bm}
\usepackage{dcolumn}
\begin{document}

\title{Local structure of ordered and disordered states of $^3$He-A in aerogel}

\author{E. V. Surovtsev}
\email{e.v.surovtsev@gmail.com}
\affiliation{P. L. Kapitza Institute for Physical Problems, Russian Academy of Science, \\
Kosygina 2, 119334 Moscow, Russia}
\affiliation{Moscow Institute of Physics and Technologies,\\
Institutskiy per. 9, Dolgoprudny, 141700, Moscow Region, Russia}

\author{I. A. Fomin}
\affiliation{P. L. Kapitza Institute for Physical Problems, Russian Academy of Science, \\
Kosygina 2, 119334 Moscow, Russia}

\date{\today}

\begin{abstract}
Random textures of the orbital part of the order parameter of
superfluid $^3$He-A in aerogel are analyzed  theoretically in the
Ginzburg and Landau region both in the presence and in the absence
of a global anisotropy. Correlation functions of angles, determining
orientation of the order parameter are found for relative distances
which are small in comparison with the characteristic scale of the
random texture. Modifications of the Larkin-Imry-Ma state in
limiting cases of a relatively strong uniaxial compression and of a
uniaxial stretching are analyzed and characteristic parameters of
the emerging states are found.
\end{abstract}

\pacs{67.30.hm, 67.30.he, 05.10.Gg, 05.40.Jc}

\maketitle

\section*{Introduction}

According to the general argument of Larkin \cite{Lark} and Imry and
Ma \cite{IMRY_Ma} (LIM) an arbitrary small quenched random field
disrupts a long range order if the order parameter is continuously
degenerate. This argument and its extension to the quenched random
anisotropy were successfully applied to impure magnetic
systems\cite{Chud1986,IMRY_1984} and other orientationally ordered
objects (for a review of the present status of the LIM effect cf.
\cite{Chud2015_1,Chud2015_2} and references therein). A character of
the resulting disordered state depends on statistical properties of
the random field and on the topology of the space of degeneracy of
the order parameter. In that sense the superfluid A-phase of liquid
$^3$He is of particular interest, it combines properties of a
superfluid with these of liquid crystals and antiferromagnets. The
order parameter of the A-phase of superfluid $^3$He with a proper
choice of the gauge can be put in the form:
\begin{equation}
A_{\mu j}=\Delta\frac{1}{\sqrt{2}}\hat d_{\mu}(\hat m_j+i\hat n_j).
\end{equation}
In the bulk liquid it is continuously degenerate with respect to
separate rotations of its spin part - unit vector $\hat d_{\mu}$ and
of the orbital part $\hat m_j+i\hat n_j$, where $\hat m_j$ and $\hat
n_j$  are two mutually orthogonal unit vectors. Usually these two
vectors are appended by the third orbital vector
$\mathbf{l}=\mathbf{m}\times\mathbf{n}$ to form an orthogonal triad.

Random anisotropy in $^3$He is produced by aerogel, immersed in the
liquid. Aerogel is a highly porous material. It is formed by
randomly oriented thin strands, their diameters (about 2-4 nm) are
much smaller than the coherence length of superfluid $^3$He $\xi_0$
and the average distance between them is of the order or bigger than
$\xi_0$\cite{parp1, Dmitriev2010}. The bulk of experimental data
indicates that orientational effect of aerogel on the orbital triad
is much stronger than on the spin vector $\hat d_{\mu}$. The latter
will be neglected in what follows.

{Volovik \cite{Volovik1996} applied the argument of LIM to the
superfluid $^3$He-A in aerogel, he} argued that the random
anisotropy induced by aerogel tends to orient locally vector
$\mathbf{l}$. This random torque disrupts the long-range order in
$^3$He-A and brings $^3$He-A into a spatially nonuniform
Larkin-Imry-Ma (LIM) state.  For silica aerogels, used in the early
experiments \cite{parp1,halp1},  both the random and a possible
global anisotropy are weak, so that in the disordered state the
order parameter preserves its form locally, but orientation of its
orbital part is different at different points.  Using general
statistical argument and $\mathbf{l}$ as the order parameter Volovik
\cite{vol2008} estimated characteristic length scale $\xi_{LIM}$ of
this state and an order of magnitude of a global anisotropy of
aerogel which would orient $\mathbf{l}$ and restore the long range
order.  These estimations agree with the experimental data
\cite{Dmitriev2010}. Nevertheless reduction of the order parameter
to one real vector $\mathbf{l}$ is not quite satisfactory. It
ignores possible effect of the ''superfluid'' degree of freedom --
rotation of $\mathbf{m}$ and $\mathbf{n}$ about the direction of
$\mathbf{l}$.  An attempt to {take this possibility into account}
was made in a previous publication of one of the present authors
\cite{fom2016}. In the Ginzburg and Landau region generally
nonlinear equations for equilibrium texture were linearized.
Linearized equations describe correctly variation of the order
parameter over a distance, which is much smaller than $\xi_{LIM}$.
To simplify the solutions and their analysis in this paper an
assumption of the absence of mass currents was introduced as a
constraint. This constraint was not physically justified. That
brings in uncertainty in quantitative results and, more important,
it does not bring in reliable  information about variation of the
''superfluid'' degree of freedom of $^3$He-A in aerogel.

In the present paper we still use the linearized equations of
equilibrium but do not impose ambiguous restrictions on their
solutions. Variation of all orbital degrees of freedom of the order
parameter is taken into account and their contribution to disruption
of the long-range order is discussed. Linearized equations are
solved analytically, their solutions contain explicit dependence on
parameters of the problem. The procedure is limited to small
variations of the order parameter but qualitative predictions about
global properties of the random textures can be obtained by
extrapolation of the found solutions.

%The paper is organized as follows:

\section{Random textures}
In the Ginzburg and Landau region contribution of interaction of
aerogel with the order parameter of superfluid $^3$He to the free
energy density can be represented in a local form \cite{fom_j}
$\eta_{jl}({\bf r})A_{\mu j}A^*_{\mu l}$, where $\eta_{jl}({\bf r})$
is a random real symmetric tensor and it varies on a distance
$\xi_c$ of the order of a distance between the strands of aerogel.
The isotropic part of interaction is included in the local
suppression of the transition temperature $T_c$ and it affects the
absolute value of the order parameter, but not its orientation. The
remaining part of $\eta_{jl}({\bf r})$ is traceless. Only this part
is kept in the following equations. For high porosity aerogels
$\eta_{jl}({\bf r})$ can be treated as a perturbation. Keeping only
orientation dependent contributions to the free energy we have:
\begin{eqnarray}
\label{F_GL} F_{GL}=N(0)\Delta^2\int d^3r\bigg\{\eta_{jl}({\bf
r})\Delta_j\Delta^*_l+\nonumber\\
\xi_s^2\left(|rot\bm{\Delta}|^2+ 3|div\bm{\Delta}|^2\right)\bigg\},
\end{eqnarray}
where $\bm{\Delta}=\frac{1}{\sqrt{2}}(\mathbf{m}+i\mathbf{n})$,
$\xi_s^2=\frac{7\zeta(3)}{12}\xi_0^2=\frac{7\zeta(3)}{12}\left(\frac{\hbar
v_F}{2\pi T_c}\right)^2$, $N(0)$ is the density of states \cite{VW}.
Variation of the functional (\ref{F_GL}) with respect to the
orientation of the triad $(\mathbf{m},\mathbf{n},\mathbf{l})$,
according to $\delta\mathbf{m}=\bm{\theta}({\bf r})\times\mathbf{m}$
etc., where $\bm{\theta}({\bf r})$ is an infinitesimal rotation
vector, renders an equation determining the equilibrium texture:
\begin{eqnarray}
\mathbf{l}\times\overrightarrow{\eta\mathbf{l}}+\xi_s^2[\mathbf{m}\times(2\nabla
(\nabla\cdot\mathbf{m})+\nabla^2\mathbf{m})+\nonumber\\
\mathbf{n}\times(2\nabla(\nabla\cdot\mathbf{n})+\nabla^2\mathbf{n})]=0.
\end{eqnarray}
{Here $\overrightarrow{\eta\mathbf{l}}$ is a vector with components
$\eta_{ij}l_j$}. Taking projections of this equation on each of the
directions $\mathbf{m},\mathbf{n},\mathbf{l}$ we arrive at three
scalar equations:
\begin{eqnarray}
\mathbf{l}\cdot(\overrightarrow{D\mathbf{m}})=\mathbf{m}\cdot(\overrightarrow{\eta\mathbf{l}}),\label{Not_1}\\
\mathbf{l}\cdot(\overrightarrow{D\mathbf{n}})=\mathbf{n}\cdot(\overrightarrow{\eta\mathbf{l}}),\label{Not_2}\\
\mathbf{n}\cdot(\overrightarrow{D\mathbf{m}})=\mathbf{m}\cdot(\overrightarrow{D\mathbf{n}})\label{Not_3},
\end{eqnarray}
where shorthand notations
$\overrightarrow{D\mathbf{m}}=\xi_s^2[2\nabla
(\nabla\cdot\mathbf{m})+\nabla^2\mathbf{m}]$ and
$\overrightarrow{D\mathbf{n}}=\xi_s^2[2\nabla
(\nabla\cdot\mathbf{n})+\nabla^2\mathbf{n}]$ are used. Solution of
these equations determines equilibrium texture for a given
realization of $\eta_{jl}({\bf r})$. Orientation of the triad
$\mathbf{m},\mathbf{n},\mathbf{l}$ is determined by  three
parameters (e.g. by the Euler angles). Derivatives of
$\mathbf{m},\mathbf{n},\mathbf{l}$, entering combinations
$\overrightarrow{D\mathbf{m}}$  and $\overrightarrow{D\mathbf{n}}$
can be expressed in terms of ``velocities'' $\omega_{a\xi}$
introduced as $\partial m_a/\partial
x_{\xi}=e_{abc}\omega_{b\xi}m_c$ etc., where $e_{abc}$ is
antisymmetric tensor and summation over repeated indices is assumed.
In these notations  Eqs. (\ref{Not_1})-(\ref{Not_3}) take the form:
\begin{eqnarray}
2(l_a n_{\xi}-n_a
l_{\xi})l_{\eta}\frac{\partial\omega_{a\xi}}{\partial
x_{\eta}}-n_a\frac{\partial\omega_{a\xi}}{\partial
x_{\xi}}+\nonumber\\
2m_b(\omega_{ba}-\omega_{ab})l_{\xi}\omega_{a\xi}+m_a\omega_{a\xi}l_b\omega_{b\xi}=\frac{l_a\eta_{ab}m_b}{\xi_s^2},\label{velocity_1}\\
2(m_a l_{\xi}-l_a
m_{\xi})l_{\eta}\frac{\partial\omega_{a\xi}}{\partial
x_{\eta}}+m_a\frac{\partial\omega_{a\xi}}{\partial
x_{\xi}}+\nonumber\\
2n_b(\omega_{ba}-\omega_{ab})l_{\xi}\omega_{a\xi}+n_a\omega_{a\xi}l_b\omega_{b\xi}=\frac{l_a\eta_{ab}n_b}{\xi_s^2},\\
2l_a\frac{\partial\omega_{a\xi}}{\partial
x_{\xi}}-l_{\xi}\frac{\partial\omega_{a\xi}}{\partial
x_a}+\omega_{a\xi}\omega_{ba}(m_b n_{\xi}-n_b
m_{\xi})=0.\label{velocity_3}
\end{eqnarray}
For  $^3$He-A projection of $\omega_{a\xi}$ on $\mathbf{l}$ is
determines the superfluid velocity:
$(v_s)_{\xi}=-\frac{\hbar}{2m}l_a\omega_{a\xi}$ and its projections
on $\mathbf{m}$ and $\mathbf{n}$ determine  $(rot
v_s)_{\xi}=-\frac{\hbar}{2m}e_{\xi\eta\zeta}(n_{a}\omega_{a\eta
})(m_{b}\omega_{b\xi})$. These relations apply to a particular
realization of texture. Definition of $\omega_{a\xi}$ includes
spatial derivatives of the order parameter. Results of averaging of
expressions containing $\omega_{a\xi}$ are sensitive to detailed
properties of the ensemble $\eta_{jl}({\bf r})$. E.g. at a formal
averaging of $\mathbf{v}_s^2$ over Gaussian ensemble of
$\eta_{ab}(\mathbf{r})$ we end up with the integral, which diverges
for large wave vectors $\mathbf{k}$. It means that the main
contribution to the ensemble average $\langle\mathbf{v}_s^2\rangle$
comes from the ''microscopic'' distances. In the present case these
are of the order of $\xi_c$ and the detailed structure of aerogel on
these distances is of importance. For a further discussion of this
question cf. Appendix A.

\section{Isotropic aerogel}
Averaged properties of textures of the order parameter depend on
statistical properties of the ensemble of tensors  $\eta_{jl}({\bf
r})$. In this section we consider a spatially isotropic ensemble,
i.e. we assume that $\langle\eta_{jl}({\bf r})\rangle=0$.  The
equilibrium texture in this case is the LIM state which can be
viewed as consisting of overlapping domains with a characteristic
size  $\xi_{LIM}$\cite{LO, IMRY_Ma, vol2008}. At distances
$R_>\gg\xi_{LIM}$ the domains are not correlated, so that the
spatial averages of vectors
$\mathbf{m}(\mathbf{r}),\mathbf{n}(\mathbf{r}),\mathbf{l}(\mathbf{r})$
over a region with a size $\sim R_>$  vanish. Random anisotropy
$\eta_{jl}(\mathbf{r})$ varies on a scale $\xi_c\ll\xi_{LIM}$. In a
window $\xi_c\ll r\ll\xi_{LIM}$ one can introduce an average
orientation of the triad $\mathbf{m}_R,\mathbf{n}_R,\mathbf{l}_R$.
Fluctuations of the orientation of the triad within the chosen
region can be expressed in terms of the small rotation vector
$\bm{\theta}({\bf r})$: $(\mathbf{m}({\bf
r})-\mathbf{m}_R)=\bm{\theta}({\bf r})\times\mathbf{m}_R$ etc..
Spatial derivatives of $\bm{\theta}({\bf r})$ render "velocities"
entering Eqs. (\ref{velocity_1})-(\ref{velocity_3}):
$\omega_{a\xi}=\partial\theta_a/\partial x_{\xi}$.  If fluctuations
are small, or if $|\bm{\theta}({\bf r})|\ll 1$, Eqs.
(\ref{velocity_1})-(\ref{velocity_3}) can be linearized over
$\bm{\theta}({\bf r})$. The linearized equations take a simple form
in a local coordinate system with axes
$\hat{\mathbf{x}},\hat{\mathbf{y}},\hat{\mathbf{z}}$ oriented along
$\mathbf{m}_R,\mathbf{n}_R,\mathbf{l}_R$ respectively (cf.
\cite{fom2016}):
\begin{eqnarray}
\nabla^2\theta_x+2\frac{\partial}{\partial z}\left(\frac{\partial
\theta_x}{\partial z}-\frac{\partial\theta_z}{\partial x}\right)
=\frac{\eta_{yz}}{\xi_s^2},
\label{linear_1}\\
\nabla^2\theta_y+2\frac{\partial}{\partial z}\left(\frac{\partial
\theta_y}{\partial z}-\frac{\partial \theta_z}{\partial y}\right)
=-\frac{\eta_{xz}}{\xi_s^2},
\\
2\nabla^2\theta_z-\frac{\partial}{\partial
z}(\nabla\cdot\bm{\theta})=0. \label{linear_3}
\end{eqnarray}
Solutions of these equations determine local properties of textures.
E.g. for two points ${\bf r}_1$ and ${\bf r}_2$ separated by a
distance $r=|{\bf r}_2-{\bf r}_1|$ meeting the condition $\xi_c\ll
r\ll\xi_{LIM}$ fluctuation of $\mathbf{l}$ is given by
$\langle(\mathbf{l}(\mathbf{r}_2)-\mathbf{l}(\mathbf{r}_1))^2\rangle=
2\langle\bm{\theta}_{\perp}(\mathbf{r}_1)\bm{\theta}_{\perp}(\mathbf{r}_1)-
\bm{\theta}_{\perp}(\mathbf{r}_2)\bm{\theta}_{\perp}(\mathbf{r}_1)\rangle
\mathbf{l}_R^2$, where
$\bm{\theta}_{\perp}(\mathbf{r}_2)\bm{\theta}_{\perp}(\mathbf{r}_1)=\bm{\theta}(\mathbf{r}_2
)\cdot\bm{\theta}(\mathbf{r}_1)-
(\bm{\theta}(\mathbf{r}_2)\cdot\mathbf{l}_R)(\bm{\theta}(\mathbf{r}_1)\cdot\mathbf{l}_R)$
i.e. $\bm{\theta}_{\perp}(\mathbf{r})$ is the projection of
$\bm{\theta}(\mathbf{r})$ on a plane, normal to $\mathbf{l}_R$.
Fluctuation of the longitudinal projection
$\theta_{\|}(\mathbf{r})=(\bm{\theta}(\mathbf{r})\cdot\mathbf{l}_R)$
can be referred shortly as a  fluctuation of the phase
$\langle(\theta_{\|}(\mathbf{r}_2)-\theta_{\|}(\mathbf{r}_1))^2\rangle=$
$2\langle\theta_{\|}(\mathbf{r}_1)\theta_{\|}(\mathbf{r}_1)-\theta_{\|}(\mathbf{r}_2)\theta_{\|}(\mathbf{r}_1)\rangle$.
It should be remarked that such terminology has direct meaning only
for small  $\bm{\theta}(\mathbf{r})$. For finite rotations
components of  $\bm{\theta}$ do not commute. Globally defined
parameters e.g. Euler angles have discontinuities when rotations
reach boundaries of the space of degeneracy of the order parameter,
that does not allow to make unambiguous separation of variation of
direction of $\mathbf{l}$ from that of the phase (for a detailed
discussion cf. \cite{VW} ch.7). The value $(\mathbf{l}_R)^2=1$ is
kept in the above formulae to preserve correct dimensionality.

Linear equations (\ref{linear_1})-(\ref{linear_3}) can be solved by
Fourier transformation:
$\theta_{x,y,z}(\mathbf{r})=\int\exp(i\mathbf{k}\mathbf{r})\theta_{x,y,z}(\mathbf{k})\frac{Vd^3k}{(2\pi)^3}$,
where $V$ is a normalization volume. Solution has more compact form
when expressed in new variables:
$\theta^{(+)}=k_x\theta_x+k_y\theta_y$,\quad
$\theta^{(-)}=k_y\theta_x-k_x\theta_y$,\quad
$\theta^{(z)}=k_z\theta_z$. Components of the tensor $\eta_{xz}$ and
$\eta_{yz}$ can be regrouped as
$\eta^{(+)}=k_x\eta_{xz}+k_y\eta_{yz}$ and
$\eta^{(-)}=k_y\eta_{xz}-k_x\eta_{yz}$, then:
\begin{eqnarray}
\theta^{(+)}(\mathbf{k})=\frac{1}{\xi_s^2k^2}f^{(+)}\eta^{(-)}(\mathbf{k}),\label{theta+}\\
\theta^{(-)}(\mathbf{k})=-\frac{1}{\xi_s^2k^2}f^{(-)}\eta^{(+)}(\mathbf{k}),\label{theta-}\\
\theta^{(z)}(\mathbf{k})=\frac{1}{\xi_s^2k^2}f^{(z)}\eta^{(-)}(\mathbf{k})\label{theta_z},
\end{eqnarray}
where $f^{(+)}=\frac{2k^2-k_z^2}{2k^2+k_z^2}$,
$f^{(-)}=\frac{k^2}{k^2+2k_z^2}$,
$f^{(z)}=\frac{k_z^2}{2k^2+k_z^2}$. In these notations
\begin{eqnarray}
\frac{\langle(\mathbf{l}(\mathbf{r_2})-\mathbf{l}(\mathbf{r_1}))^2\rangle}{\mathbf{l}_R^2}=2\int[1-\exp(i\mathbf{k}\cdot\mathbf{r})]\times\nonumber\\
\frac{\langle\theta^{(+)}(-\mathbf{k})\theta^{(+)}(\mathbf{k})+\theta^{(-)}(-\mathbf{k})\theta^{(-)}(\mathbf{k})\rangle}{k_z^2-k^2}
\frac{Vd^3k}{(2\pi)^3},\label{cor_l_isotropic}
\end{eqnarray}
where $\mathbf{r}=\mathbf{r}_2-\mathbf{r}_1$. Using expressions
(\ref{theta+}) and (\ref{theta-}) we conclude that the principal
contribution to the integral comes from the region of $k\sim 1/r$.
In what follows distances $r\gg\xi_c$, or wave-vectors $k\ll
1/\xi_c$ are of interest. Strands of silica aerogel are correlated
on a distance $\sim \xi_c$ \cite{parp1, halp1} and for the actual
values of $k$ aerogel can be considered as ensemble of
non-correlated impurities. Then the correlation functions, entering
expression in Eq. (\ref{cor_l_isotropic}) do not depend on $k$:
$\langle\eta_{xz}(-\mathbf{k})\eta_{xz}(\mathbf{k})\rangle=\langle\eta_{yz}(-\mathbf{k})\eta_{yz}(\mathbf{k})\rangle=const.\equiv
K/V$ and
$\langle\eta_{xz}(-\mathbf{k})\eta_{yz}(\mathbf{k})\rangle=0$. With
these assumptions
\begin{eqnarray}
\frac{\langle(\mathbf{l}(\mathbf{r}_2)-\mathbf{l}(\mathbf{r}_1))^2\rangle}{\mathbf{l}_R^2}=2\int[1-\exp(i\mathbf{k}\cdot\mathbf{r})]
\frac{K}{\xi_s^4k^4}\times\nonumber\\
\left[(f^{(+)})^2+(f^{(-)})^2\right]\frac{d^3k}{(2\pi)^3}.\label{cor_l_isotropic_1}
\end{eqnarray}
For the fluctuation of phase analogous argument renders:
\begin{eqnarray}
\langle(\theta_{\|}(\mathbf{r}_2)-\theta_{\|}(\mathbf{r}_1))^2\rangle=2\int[1-\exp(i\mathbf{k}
\cdot\mathbf{r})]\frac{K}{\xi_s^4k^4}\times\nonumber\\
\frac{k^2-k_z^2}{k_z^2}(f^{(z)})^2\frac{d^3k}{(2\pi)^3}.~~\label{cor_theta_parallel}
\end{eqnarray}
Analysis of dimensions shows that integrals in the r.h.s of Eq.
(\ref{cor_l_isotropic_1}) and (\ref{cor_theta_parallel}) are
proportional to $r$, as it has to be at a random walk. The
coefficients can be represented as $\frac{A_{\perp,\|}}{\xi_{LIM}}$,
where $\xi_{LIM}=\frac{2\pi\xi_s^4}{K}$ is the characteristic length
expressed in terms of parameters of the problem and $A_{\perp,\|}$
are  coefficients of the order of unity. They depend on orientation
of $\mathbf{r}$ with respect to $\mathbf{l}$. Analytical expressions
for general orientation are cumbersome, they are presented in
Appendix B. Here we quote results only for $\mathbf{r}\|\mathbf{l}$
and $\mathbf{r}\perp\mathbf{l}$ :

1) for $\mathbf{r}\|\mathbf{l}$
\begin{eqnarray}
\frac{\langle(\mathbf{l}(\mathbf{r_2})-\mathbf{l}(\mathbf{r_1}))^2\rangle}{\mathbf{l}_R^2}=\frac{r}{\xi_{LIM}}\left(2-4\ln\frac{3}{2}\right)
\nonumber\\
\approx0.38\cdot\frac{r}{\xi_{LIM}}~~~\label{asymp_l_par}\\
\langle(\theta_{\|}(\mathbf{r}_2)-\theta_{\|}(\mathbf{r}_1))^2\rangle=\frac{r}{\xi_{LIM}}\left(\frac{5}{2}\ln\frac{3}{2}-1\right)
\nonumber\\
\approx0.014\cdot\frac{r}{\xi_{LIM}}~~~\label{asymp_theta_par}
\end{eqnarray}

2) for $\mathbf{r}\perp\mathbf{l}$
\begin{eqnarray}
\frac{\langle(\mathbf{l}(\mathbf{r_2})-\mathbf{l}(\mathbf{r_1}))^2\rangle}{\mathbf{l}_R^2}=\frac{r}{\xi_{LIM}}\left(\frac{51+\sqrt{3}-20\sqrt{6}}{6}\right)
\nonumber\\
\approx0.62\cdot\frac{r}{\xi_{LIM}}~~~\label{asymp_l_per}\\
\langle(\theta_{\|}(\mathbf{r}_2)-\theta_{\|}(\mathbf{r}_1))^2\rangle=\frac{r}{\xi_{LIM}}\left(\frac{9\sqrt{6}-22}{4}\right)\nonumber\\
\approx0.011\cdot\frac{r}{\xi_{LIM}}~~~\label{asymp_theta_per}
\end{eqnarray}
At $r\sim\xi_{LIM}$ fluctuations are of the order of unity and the
long range order is disrupted. For both considered orientations
within the limits of applicability of linear approximation the rate
of change of the orientation of $\mathbf{l}$ is significantly
greater than the rate of de-phasing so that the disruption of the
long-range order is mainly due to the random variation of
orientation of $\mathbf{l}$. That explains why the previously
imposed restrictions \cite{vol2008,fom2016} do not effect
significantly estimations of the characteristic length. The
relatively small rate of change of the "phase" correlator may be due
to the absence of direct coupling of $\theta_z$ to the random
anisotropy.

Numerical estimation of $\xi_{LIM}$ can be made with the aid of the
"Model of Random Cylinders" (MRC) \cite{thuneb_1998,vol2008,
SF2008}. {In this model aerogel is assumed to consist of cylinders
of the same radius $r_a$ and height $h$. Tensor
$\eta_{jl}(\mathbf{r})$ can be found using theory of Rainer and
Vuourio\cite{Rainer} of small objects in superfluid $^3$He. The
smallness of an object is controlled by the condition
$\sigma_{tr}/\xi_0^2\ll 1$, where $\sigma_{tr}$ is transport
cross-section of a single impurity. The average value of tensor
$\eta_{jl}(\mathbf{r})$ and the value of $K(\mathbf{k})$ are
proportional to $n\xi_0\sigma_{tr}$ and $n(\xi_0\sigma_{tr})^2$
respectively. Here $n$ is concentration of impurities. Since for
cylinder $\sigma_{tr}\sim r_ah$ it can be shown that $K$ is
proportional to $\xi_0^2h(1-P)$ or to be more precise\cite{fom2016}
$K=\frac{3\pi^5}{5\cdot2^{10}}\xi_0^2h(1-P)$, where P is porosity of
aerogel. Following our definition of $\xi_{LIM}$ one can find that
$\xi_{LIM}\approx 7\frac{\xi_0^2}{h\cdot(1-P)}$ and if we take
$h\sim\xi_c\sim\xi_0$ and $P\sim 0.98$, then $\xi_{LIM}\sim10$
$\mu$m for high pressures. It agrees with the estimations of
Thuneberg \cite{thuneb_1998}. It should be noted, that the condition
of applicability of Rainer and Vuorio theory is fulfilled under our
assumptions since $\sigma_{tr}/\xi_0^2\sim r_ah/\xi_0^2\sim
r_a/\xi_0\ll1$.}

The long range order and property of superfluidity can be restored
if continuous degeneracy of the order parameter over orientation  of
$\mathbf{l}$ is lifted e.g. by a global anisotropy.

\section{Global anisotropy}
As prepared samples of aerogel can have appreciable macroscopic
anisotropy. In a controlled way global anisotropy can be produced by
deformation of an originally isotropic sample \cite{knmts, FS2015}.
Formally global anisotropy is described by an extra term in the
energy functional (\ref{F_GL}) $\kappa_{jl}\Delta_j\Delta_l^*$,
where $\kappa_{jl}$ is a uniform symmetric traceless real tensor.
The ensuing modification of the equations of equilibrium consists in
substitution of combination $\eta_{jl}(\mathbf{r})+\kappa_{jl}$
instead of the $\eta_{jl}(\mathbf{r})$, so that the notation
$\eta_{jl}$ is preserved for purely random anisotropy. Eq.
(\ref{Not_3}) does not change and Eqs. (\ref{Not_1}) and
(\ref{Not_2}) acquire additional terms, depending on $\kappa_{jl}$:

\begin{eqnarray}
\mathbf{l}\cdot(\overrightarrow{D\mathbf{m}})=\mathbf{m}\cdot(\overrightarrow{\eta\mathbf{l}})+
\mathbf{m}\cdot(\overrightarrow{\kappa\mathbf{l}}),
\label{Not_11}     \\
\mathbf{l}\cdot(\overrightarrow{D\mathbf{n}})=\mathbf{n}\cdot(\overrightarrow{\eta\mathbf{l}})+
\mathbf{n}\cdot(\overrightarrow{\kappa\mathbf{l}}).\label{Not_21}
\end{eqnarray}
We start from the situation when global anisotropy is much stronger
than random anisotropy. It is convenient to introduce except for the
"moving" coordinate system $x,y,z$ a "static" one with the axes
$u,v,w$ oriented along the principal directions of $\kappa_{jl}$. In
zero order approximation over $\eta_{jl}(\mathbf{r})$ Eqs.
(\ref{Not_11}) and (\ref{Not_21}) have spatially uniform solution
meeting the conditions:
$\mathbf{m}\cdot(\overrightarrow{\kappa\mathbf{l}})=0$ and
$\mathbf{n}\cdot(\overrightarrow{\kappa\mathbf{l}})=0$. It means
that $\mathbf{l}$ is oriented along one of the principal directions
of $\kappa_{jl}$. The lowest free energy corresponds to the largest
principal value of the three $\kappa_u,\kappa_v,\kappa_w$. We
consider here axially symmetric global anisotropy. In this case all
three principal values can be expressed in terms of one parameter
$\kappa=\kappa_u=\kappa_v$ and $\kappa_w=-2\kappa$. Orientation of
the triad $\mathbf{m},\mathbf{n},\mathbf{l}$ with respect to the
axis of anisotropy depends on a sign of $\kappa$.

\subsection{Uniform compression} 
\begin{figure}[h]
\includegraphics[scale=0.45]{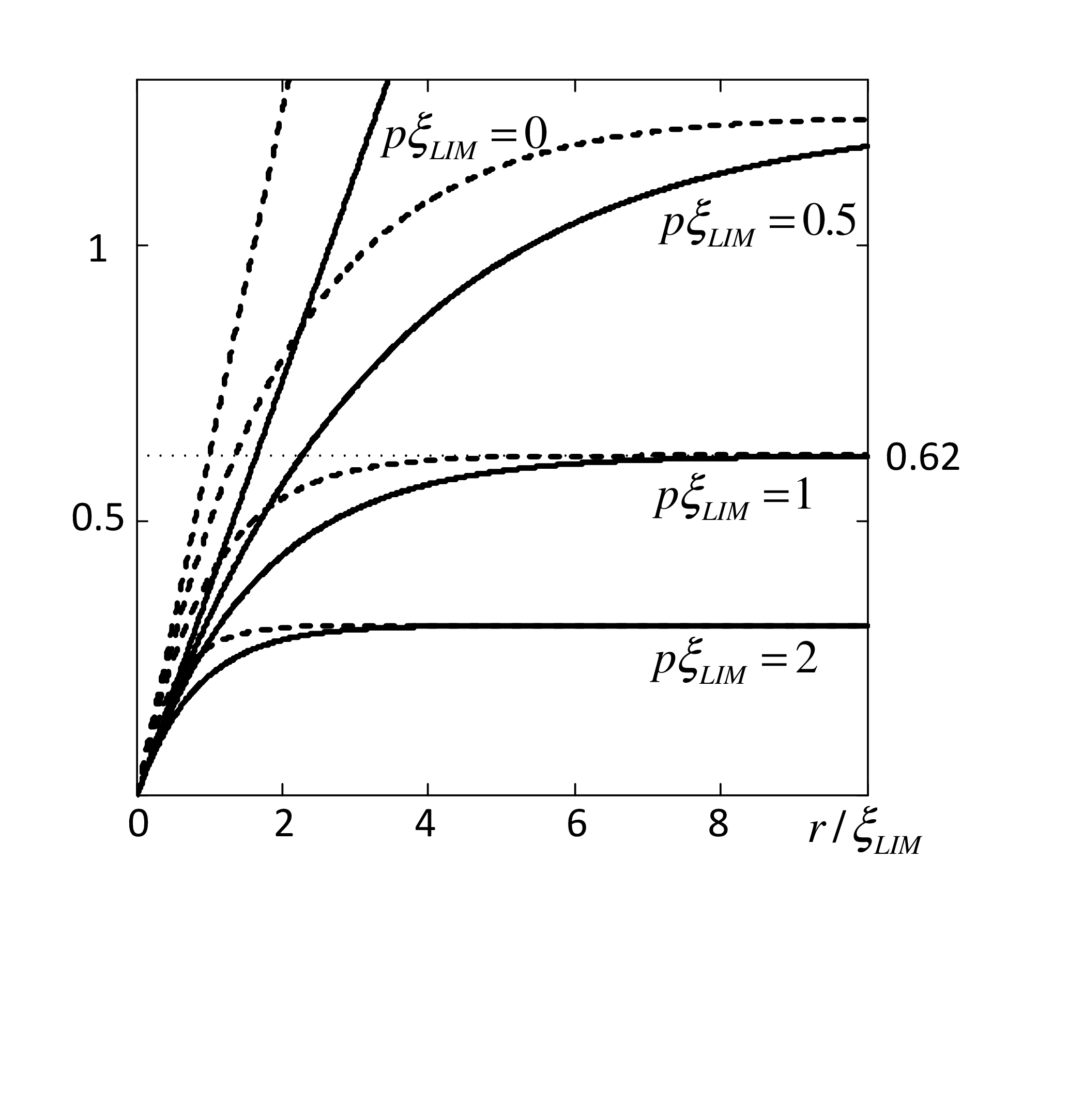}
\caption{\label{fig:epsart1} Dependence of
${\langle(\mathbf{l}(\mathbf{r}_2)-\mathbf{l}(\mathbf{r}_1))^2\rangle}/{\mathbf{l}_R^2}$
on $r/\xi_{LIM}$ for uniform compression for four values of
parameter $p\xi_{LIM}=0,~0.5,~1,~2$. Solid lines correspond to the
direction $\mathbf{r}\|\mathbf{w}$ and dashed lines are for
$\mathbf{r}\perp\mathbf{w}$. Straight lines show asymptotical
dependence of correlator when $p\xi_{LIM}\longrightarrow0$, Eqs.
(\ref{asymp_l_par}) and (\ref{asymp_theta_par}). Horizontal line
demonstrates asymptotical value of correlator from Eq.
(\ref{asymp1})  when $p\xi_{LIM}=1$. }
\end{figure}

\begin{figure}[h]
\includegraphics[scale=0.45]{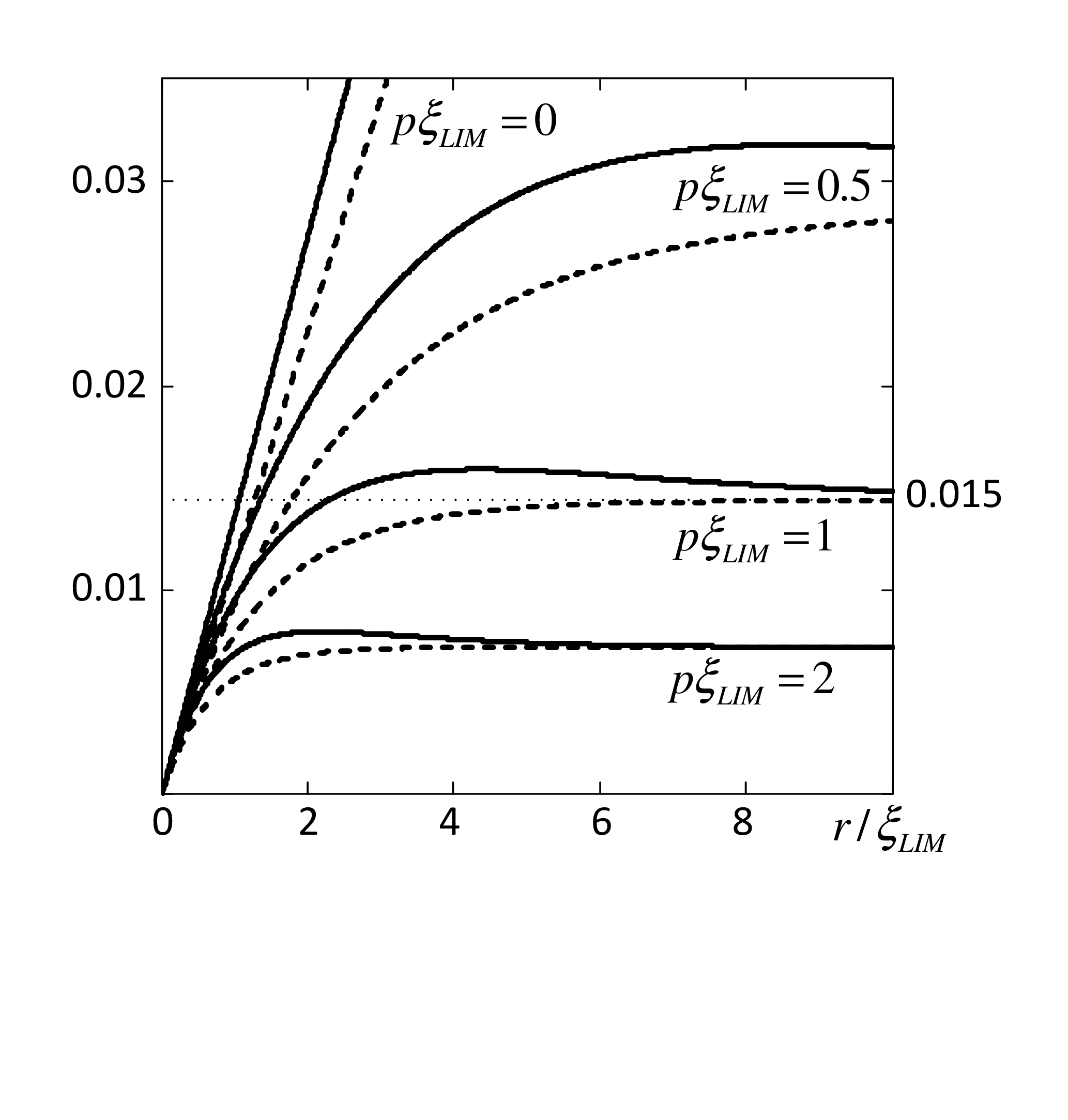}
\caption{\label{fig:epsart2} Dependence of
$\langle(\theta_{\|}(\mathbf{r}_2)-\theta_{\|}(\mathbf{r}_1))^2\rangle$
on $r/\xi_{LIM}$ for uniform compression for four values of
parameter $p\xi_{LIM}=0,~0.5,~1,~2$. Solid lines correspond to the
direction $\mathbf{r}\|\mathbf{w}$ and dashed lines are for
$\mathbf{r}\perp\mathbf{w}$. Straight lines show asymptotical
dependence of correlator when $p\xi_{LIM}\longrightarrow0$, Eqs.
(\ref{asymp_l_per}) and (\ref{asymp_theta_per}). Horizontal line
demonstrates asymptotical value of correlator from Eq.
(\ref{asymp2})  when $p\xi_{LIM}=1$. }
\end{figure}

At  a uniform uniaxial compression $\kappa<0$ \cite{Dmitriev2010}. Zero order $\mathbf{l}$
is aligned or counter-aligned with the axis $\mathbf{w}$. Both
states have the same energy, so they can coexist as domains. Energy
of the domain wall is positive, in the equilibrium a one-domain
state is favored and the long-range order exists. Random anisotropy
induces small deviations of the order parameter from its equilibrium
orientation. These deviations can be expressed in terms of a small
vector  $\bm{\theta}$, which is defined as in the isotropic case via
$\delta\mathbf{m}(\mathbf{r})=\bm{\theta}({\bf r})\times\mathbf{m}$.
It is convenient to choose coordinate axes $x,y,z$ so that
$\hat{\mathbf{z}}$ is aligned with the axis of anisotropy (and with
equilibrium direction of $\mathbf{l}$) and
$\hat{\mathbf{x}},\hat{\mathbf{y}}$ are directed along equilibrium
orientations $\mathbf{m},\mathbf{n}$ respectively. In these
notations the linearized equations (\ref{Not_11}), (\ref{Not_21})
and (\ref{Not_3})  acquire the following form:
\begin{eqnarray}
\nabla^2\theta_x+p^2\theta_x+2\frac{\partial}{\partial
z}\left(\frac{\partial \theta_x}{\partial
z}-\frac{\partial\theta_z}{\partial x}\right)
=\frac{\eta_{yz}}{\xi_s^2},\label{linear_1c}
\\
\nabla^2\theta_y+p^2\theta_y+2\frac{\partial}{\partial
z}\left(\frac{\partial \theta_y}{\partial z}-\frac{\partial
\theta_z}{\partial y}\right) =-\frac{\eta_{xz}}{\xi_s^2},\\
2\nabla^2\theta_z-\frac{\partial}{\partial
z}(\nabla\cdot\bm{\theta})=0,\label{linear_3c}
\end{eqnarray}
where $p^2=3|\kappa|/\xi^2_s$. An argument, analogous to that of the
previous section renders:
\begin{eqnarray}
\theta^{(+)}(\mathbf{k})=f^{(+)}\frac{\bar{\eta}^{(-)}(\mathbf{k})}{\xi_s^2(k^2+p^2f^{(+)})},\label{theta+compress}
\\
\theta^{(-)}(\mathbf{k})=-f^{(-)}\frac{\bar{\eta}^{(+)}(\mathbf{k})}{\xi_s^2(k^2+p^2f^{(-)})}\label{theta-compress},
\\
\theta^{(z)}(\mathbf{k})=f^{(z)}\frac{\bar{\eta}^{(-)}(\mathbf{k})}{\xi_s^2(k^2+p^2f^{(+)})}.
\end{eqnarray}
Substitution of expressions (\ref{theta+compress}) and
(\ref{theta-compress}) in Eq. (\ref{cor_l_isotropic}) renders
\begin{eqnarray}
\label{Compress_l}
\frac{\langle(\mathbf{l}(\mathbf{r_2})-\mathbf{l}(\mathbf{r_1}))^2\rangle}{\mathbf{l}_R^2}=2\int\frac{Kd^3k}
{(2\pi)^3\xi_s^4}[1-\exp(i\mathbf{k}\cdot\mathbf{r})]\times\nonumber\\
\left\{\frac{(f^{(-)})^2}{(k^2+p^2f^{(-)})^2}+\frac{(f^{(+)})^2}{(k^2+p^2f^{(+)})^2}\right\}.~~
\end{eqnarray}
This expression contains an extra parameter of length: $1/p$. The
integral in the r.h.s. has qualitatively different asymptotic
behavior at $r\gg 1/p$ and $r\ll 1/p$, that can be easily
demostrated for a particular case  $\mathbf{r}\|\mathbf{w}$.  Using
$u=\cos(\mathbf{k}\mathbf{w})$ and $k=|\mathbf{k}|$ as independent
variables and integrating over $dk$ we can rewrite Eq.
(\ref{Compress_l}) in the following form:
\begin{eqnarray}
\frac{\langle(\mathbf{l}(\mathbf{r}_2)-\mathbf{l}(\mathbf{r}_1))^2\rangle}{\mathbf{l}_R^2}=\frac{r}{\xi_{LIM}}\int^1_0
du\frac{u}{2}\times
\nonumber\\
\left\{\left[\exp(-q^{(-)})+\frac{1}{q^{(-)}}(1-\exp(-q^{(-)}))\right](f^{(-)})^2+\right.\nonumber\\
\left.\left[\exp(-q^{(+)})+\frac{1}{q^{(+)}}(1-\exp(-q^{(+)}))\right](f^{(+)})^2\right\},~\label{cor_l_compression}
\end{eqnarray}
where $q^{(\pm)}=pru\sqrt{f^{(\pm)}}$. The functions $f^{(-)}$ and
$f^{(+)}$ are limited and have no singularities in the interval
$0<u<1$. At $pr\rightarrow 0$ each of the square brackets in the
integral in Eq. (\ref{cor_l_compression}) tends to 2 and we recover
the result for isotropic aerogel, as it was expected. In the
opposite limit $pr\rightarrow\infty$ exponents in the integral in
Eq. (\ref{cor_l_compression}) can be omitted and the leading term in
the asymptotic does not depend on $r$:
\begin{eqnarray}
\frac{\langle(\mathbf{l}(\mathbf{r}_2)-\mathbf{l}(\mathbf{r}_1))^2\rangle}{\mathbf{l}_R^2}\rightarrow\int^1_0
\frac{du}{2}
\frac{\left(f^{(+)}\right)^{\frac{3}{2}}+\left(f^{(-)}\right)^{\frac{3}{2}}}{p\xi_{LIM}}\nonumber
\\
=\frac{u}{2}\int^1_0 \frac{du}{2p\xi_{LIM}}
\left[\left(\frac{2-u^2}{2+u^2}\right)^{\frac{3}{2}}+\frac{1}{(1+2u^2)^{\frac{3}{2}}}\right]\approx
\frac{0.62}{p\xi_{LIM}}.\nonumber \\
 \label{asymp1}
\end{eqnarray}
This limiting value is valid for any direction of $\mathbf{r}$ with
respect to $\mathbf{w}$ (FIG.~\ref{fig:epsart1}). In the limit
$pr\gg 1$ one can neglect $\exp(i\mathbf{k}\cdot\mathbf{r})$ under
the integral sign in Eq. (\ref{Compress_l}) because of its fast
oscillations.
If $p\xi_{LIM}\gg1$ fluctuations of $\mathbf{l}(\mathbf{r})$ remain
small for all distances $r=|\mathbf{r}_2-\mathbf{r}_1|$ and
orientation of $\mathbf{l}$ varies within a narrow cone. In this
situation $\theta_{\|}(\mathbf{r})$ can be considered as the phase
of the order parameter. Application of the above argument to
fluctuation of phase renders:
\begin{eqnarray}
\langle(\theta_{\|}(\mathbf{r}_2)-\theta_{\|}(\mathbf{r}_1))^2\rangle=2\int\frac{Kd^3k}
{(2\pi)^3\xi_s^4}
[1-\exp(i\mathbf{k}\cdot\mathbf{r})]\times\nonumber
\\
\frac{(k^2-k_z^2)}{k_z^2} \frac{(f^{(z)})^2}{(k^2+p^2f^{(+)})^2}.~~~
\end{eqnarray}
For $\mathbf{r}\|\mathbf{w}$ transformation, analogous to that
preceding Eq. (\ref{cor_l_compression}) renders
\begin{eqnarray}
\langle(\theta_{\|}(\mathbf{r}_2)-\theta_{\|}(\mathbf{r}_1))^2\rangle=\frac{r
K}{4\pi\xi_s^4}\int^1_0 u du
\frac{u^2(1-u^2)}{2+u^2}\times\nonumber\\
\left[\exp(-q^{(+)})+\frac{1}{q^{(+)}}(1-\exp(-q^{(+)})\right].~~
\end{eqnarray}
This expression has different asymptotic behavior at small and large
distances, FIG.~\ref{fig:epsart2}.
At $pr\ll 1$ square of fluctuation of $\theta_{\|}$ grows linearly
with $r$ but the rate of growth is very small:
\begin{eqnarray}
\langle(\theta_{\|}(\mathbf{r}_2)-\theta_{\|}(\mathbf{r}_1))^2\rangle\approx\frac{r}{\xi_{LIM}}\left(\frac{5}{2}\ln\frac{3}{2}-1\right).
\end{eqnarray}
In the opposite limit $pr\to\infty$  the  fluctuation tends to a
constant
\begin{eqnarray}
\langle(\theta_{\|}(\mathbf{r}_2)-\theta_{\|}(\mathbf{r}_1))^2\rangle\rightarrow\nonumber
\\
\frac{1}{p\xi_{LIM}}\int^1_0 \frac{du}{2}
\frac{u^2(1-u^2)}{(2+u^2)\sqrt{(4-u^2)}}\approx
0.015\cdot\frac{1}{p\xi_{LIM}}.~~~\label{asymp2}
\end{eqnarray}

At sufficiently large $p\xi_{LIM}$ fluctuation is small and the
long-range order is preserved at least within one domain. The
one-domain state is the true equilibrium state. Because of the
pinning of the domain walls by fluctuations of random anisotropy
meta-stable multi-domain states can be realized as well. In this
limit they consists of well defined domains, separated by the domain
walls with a width  $\sim 1/p$.

When anisotropy is getting weaker the energetic advantage of the
ordered state in comparison with the disordered decreases. In a
region $p\xi_{LIM}\sim 1$ free energies of two states become equal
and they interchange their roles via a first order phase transition.
Because of a pinning of domain walls hysteresis phenomena are
expected and structure of concrete state depends on a history of its
preparation.

In the ordered state small local fluctuations of orientation of
$\mathbf{l}$ effect directly the value of c.w. NMR shift
\cite{Dmitriev2010}. The shift is proportional to
$Q=\frac{3}{2}(\langle l_w^2\rangle-\frac{1}{3})$. With the use of
previous calculations
\begin{eqnarray}
\label{NMR_shift} \langle
l_w^2\rangle=1-\frac{1}{2p\xi_{LIM}}\int^1_0 \frac{du}{2}
\left[(f^{(-)})^{3/2}+(f^{(+)})^{3/2}\right]. ~~~
\end{eqnarray}
For the moment we don`t know of a systematic experimental study of
this effect.

\subsection{Uniform stretching} A positive $\kappa$ is realized
when aerogel is uniaxially stretched. In real experiments because of
fragility of aerogel a state with $\kappa>0$ is prepared by axially
symmetric compression of a cylindrical sample in directions
perpendicular to its symmetry axis \cite{Elbs,Halperin_orient}. A
favorite orientation of $\mathbf{l}$ in this case is any direction
perpendicular to the symmetry axis $\mathbf{w}$
($\mathbf{l}=(l_u,l_v,0))$. We can choose coordinate so that $l_u=0$
at the point of observation, then equations for small fluctuations
of orientation of the triad $\mathbf{m},\mathbf{n},\mathbf{l}$ are
analogous to the equations (\ref{linear_1c})-(\ref{linear_3c}) with
obvious changes:
\begin{eqnarray}
\nabla^2\theta_w+2\frac{\partial}{\partial
v}\left(\frac{\partial\theta_w}{\partial v}-\frac{\partial
\theta_v}{\partial
w}\right)=\frac{\eta_{uv}}{\xi_s^2},\label{linear_1s}
\\
\nabla^2\theta_u-p^2\theta_u+2\frac{\partial}{\partial
v}\left(\frac{\partial\theta_u}{\partial v}-\frac{\partial
\theta_v}{\partial u}\right)=-\frac{\eta_{vw}}{\xi_s^2},
\\
2\nabla^2\theta_v-\frac{\partial}{\partial
v}(\nabla\cdot\bm{\theta})=0.\label{linear_3s}
\end{eqnarray}
\begin{figure}[h]
\includegraphics[scale=0.45]{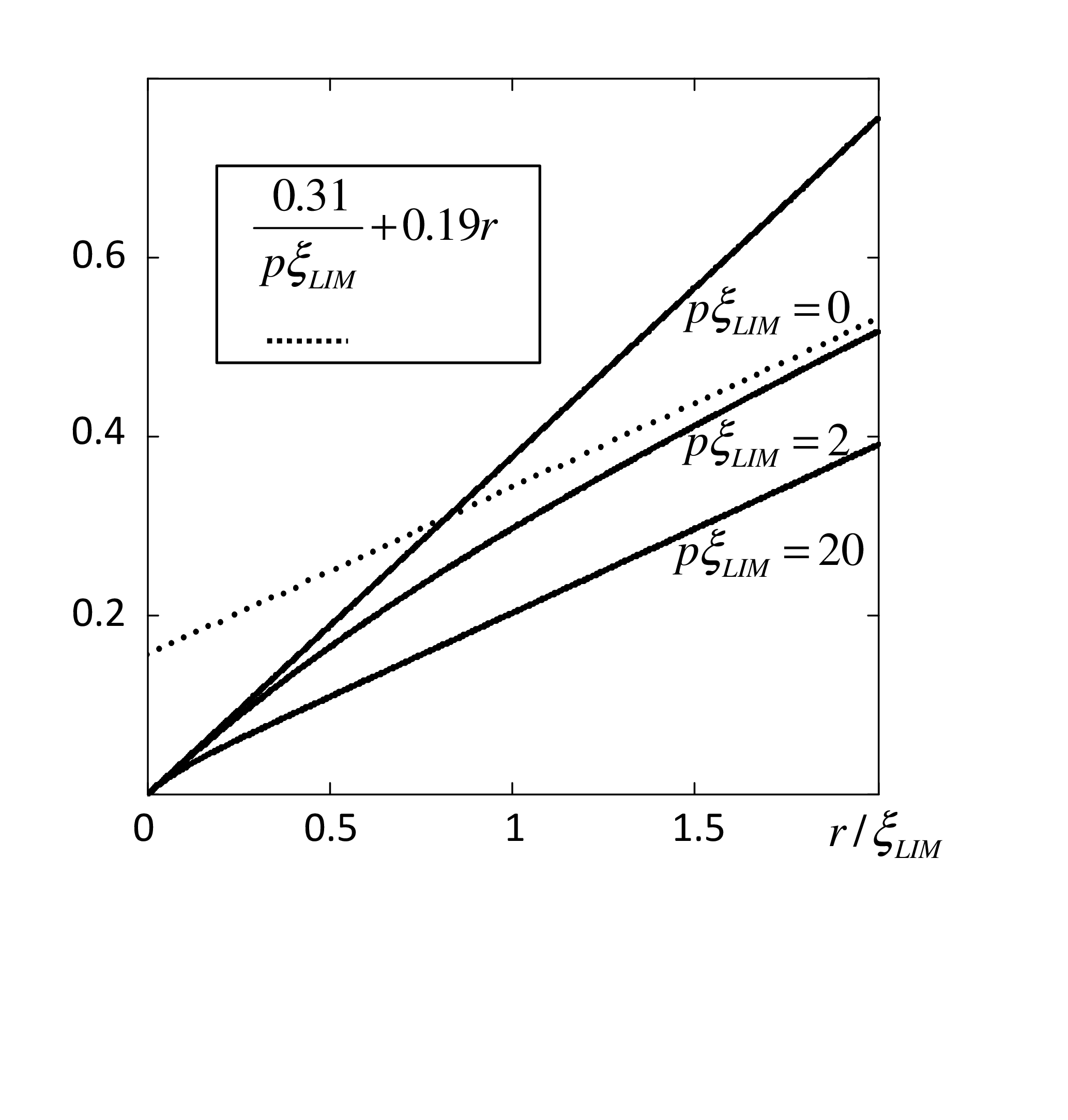}
\caption{\label{fig:epsart3} Dependence of
${\langle(\mathbf{l}(\mathbf{r}_2)-\mathbf{l}(\mathbf{r}_1))^2\rangle}/{\mathbf{l}_R^2}$
on $r/\xi_{LIM}$ for uniform stretching for three values of
parameter $p\xi_{LIM}=0,~2,~20$ and $\mathbf{r}\|\mathbf{w}$. Dotted
line shows asymptotical dependence of correlator when $r\cdot
p\longrightarrow \infty$. The upper solid line ($p\xi_{LIM}=0$) is
taken from Eq.~\ref{asymp_l_per}.}
\end{figure}
\begin{figure}[h]
\includegraphics[scale=0.45]{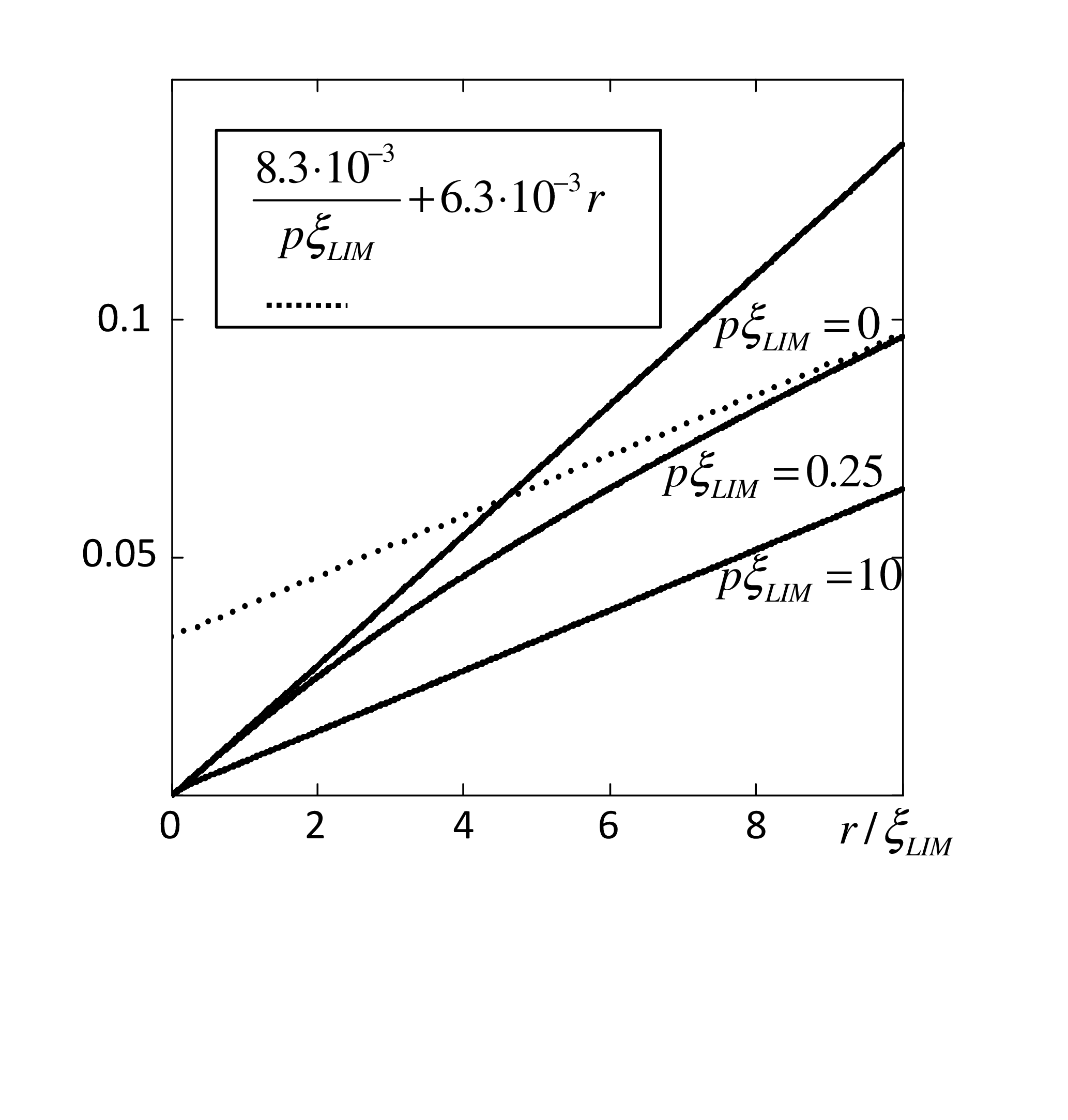}
\caption{\label{fig:epsart4} Dependence of
$\langle(\theta_{\|}(\mathbf{r}_2)-\theta_{\|}(\mathbf{r}_1))^2\rangle$
on $r/\xi_{LIM}$ for uniform stretching for three values of
parameter $p\xi_{LIM}=0,~0.25,~10$ and $\mathbf{r}\|\mathbf{w}$.
Dotted line show asymptotical dependence of correlator when $r\cdot
p\longrightarrow \infty$. The upper solid line ($p\xi_{LIM}=0$) is
taken from Eq.  \ref{asymp_theta_per}.}
\end{figure}
Solution of the equations (\ref{linear_1s})-(\ref{linear_3s})
follows the same line as for the case $\kappa<0$. Essential
difference is that now only one degree of freedom remains "gapped",
it is rotation $\theta_u$, which takes $\mathbf{l}$ out of the $u,v$
plane. Two other rotations $\theta_v$ and $\theta_w$   move the
triad $\mathbf{m},\mathbf{n},\mathbf{l}$ within its space of
degeneracy. At a strong anisotropy $\mathbf{l}$ moves within the
$u,v$ plane, but has random orientation within this plane. This
state is referred as 2D LIM state\cite{Dmitriev2010}.
\begin{figure}[h]
\includegraphics[scale=0.45]{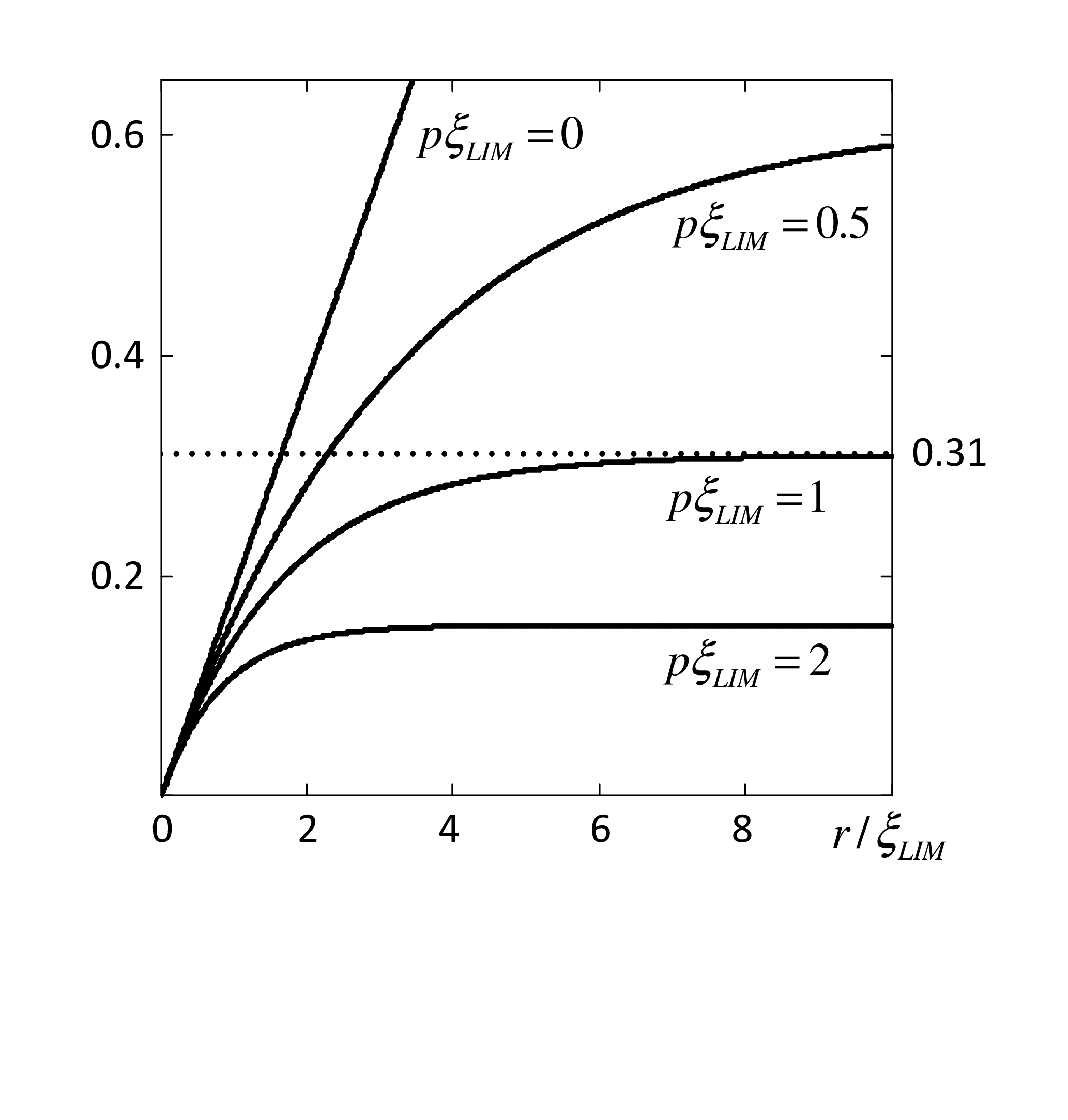}
\caption{\label{fig:epsart5} Dependence of
$\langle(\theta_{u}(\mathbf{r}_2)-\theta_{u}(\mathbf{r}_1))^2\rangle$
on $r/\xi_{LIM}$ for uniform stretching for four values of parameter
$p\xi_{LIM}=0,~0.5,~1,~2$ and $\mathbf{r}\|\mathbf{w}$. Dotted line
show asymptotical dependence of correlator when $r\cdot
p\longrightarrow \infty$.}
\end{figure}

%The solution of the above equations takes a form of multiple
%integrals.
The expressions for the correlators take a form of multiple
integrals. We present here only results of numerical integration for
the case when $\mathbf{r}\|\mathbf{w}$ (FIG.3-5). Full expressions
for integrals are given in Appendix C. In the region of ${r}\ll1/p$
the dependence of correlators on $\mathbf{r}$ is linear, for
${\langle(\mathbf{l}(\mathbf{r}_2)-\mathbf{l}(\mathbf{r}_1))^2\rangle}/{\mathbf{l}_R^2}$
and
$\langle(\theta_{\|}(\mathbf{r}_2)-\theta_{\|}(\mathbf{r}_1))^2\rangle$
it is given by Eqs. (\ref{asymp_l_par}) and (\ref{asymp_theta_par}).
The dependence of correlator
$\langle(\theta_{u}(\mathbf{r}_2)-\theta_{u}(\mathbf{r}_1))^2\rangle$
in this region is close to
$\frac{1}{2}{\langle(\mathbf{l}(\mathbf{r}_2)-\mathbf{l}(\mathbf{r}_1))^2\rangle}/{\mathbf{l}_R^2}$.
In the opposite limit ${r}\gg1/p$ dependencies of the first two
correlators can be approximated by the linear function
$\frac{a}{p\xi_{LIM}}+b\cdot r$. The values of the coefficients $a$
and $b$ are given on the inserts on FIG.~\ref{fig:epsart3} and
FIG.~\ref{fig:epsart4}. The limiting value of
$\langle(\theta_{u}(\mathbf{r}_2)-\theta_{u}(\mathbf{r}_1))^2\rangle$
at $rp\gg1$ defines $\langle
l_{w}^2\rangle=\frac{1}{2}\langle(\theta_{u}(\mathbf{r}_2)-\theta_{u}(\mathbf{r}_1))^2\rangle$
in linear regime, i.e. if $p\xi_{LIM}\gg1$. It is found to be
$\frac{0.31}{p\xi_{LIM}}$, that is approximately half of the
limiting value of correlator
${\langle(\mathbf{l}(\mathbf{r}_2)-\mathbf{l}(\mathbf{r}_1))^2\rangle}/{\mathbf{l}_R^2}$
for the case of uniform compression (Eq. (\ref{asymp1})).

\section{Discussion}
The approach, based on linearization of the equations of equilibrium
is complementary to that, based on the argument of LIM. Linearized
equations render a "zoomed" picture of a small region within the LIM
state or within the similar random textures of the order parameter
of $^3$He-A in aerogel. Solutions of these equations present a
quantitative description of a decrease of correlations of
orientation of the order parameter and development of disorder with
an increase of a distance  between the two points within the chosen
region.  They describe also recovery of the long-range order when
sufficiently strong global anisotropy is applied.  In comparison
with magnetic glasses the order parameter of  $^3$He-A has
additional degree of freedom -- "phase" variable. This variable is
getting disordered together with the vector $\mathbf{l}$ although,
unlike $\mathbf{l}$, the "phase" variable does not couple directly
to the random anisotropy. Extrapolation of the results of linear
analysis to distances of the order of  $\xi_{LIM}$ matches the
results based on the LIM argument. A qualitative picture obtained by
such matching can be used as a guidance for a further quantitative
description of random textures including the region
$r\sim\xi_{LIM}$, where nonlinearities become significant. That
requires a serious numerical work, but it would render important
results, e.g. a precise value of the critical global anisotropy, at
which the phase transition from the disordered to the ordered state
occurs, extension of the result for the NMR shift (Eq.
(\ref{NMR_shift})) in a region of finite fluctuations of the order
parameter and description of global properties of textures, which
can not be analyzed within the linear approximation.

\section{Acknowledgements}
We thank V.V.Dmitriev for useful discussions and comments. This work
was supported in part by the Russian Foundation for Basic Research,
project \# 14-02-00054-a and the Basic Research Program of the
Presidium of Russian Academy of Sciences

\appendix

\section{}
In the linear approximation the superfluid velocity is determined by
the gradient of $\theta_{\|}(\mathbf{r})$:
$(v_s)_{\xi}=-\frac{\hbar}{2m}\frac{\partial\theta_{\|}}{\partial
x_{\xi}}$. We are interested in the ensemble average
\begin{equation}
\langle v_s^2\rangle=\left(\frac{\hbar}{2m}\right)^2\left\langle
\left(\frac{\partial\theta_{\|}}{\partial\textbf{r}}\right)^2\right\rangle,
\end{equation}
or in terms of Fourier components:
\begin{equation}
\langle v_s^2\rangle=\left(\frac{\hbar}{2m}\right)^2\int
k^2\langle\theta_{\|}(\mathbf{k})\theta_{\|}(-\mathbf{k})\rangle
\frac{Vd^3k}{(2\pi)^3}
\end{equation}
with the aid of Eq. (15) $\langle v_s^2\rangle$ can be expressed in
terms of correlation functions of $\eta_{jl}(\mathbf{k})$. We assume
that aerogel is globally isotropic, then
$\langle\eta_{xz}(-\mathbf{k})\eta_{yz}(\mathbf{k})\rangle=0$ and
$\langle\eta_{xz}(-\mathbf{k})\eta_{xz}(\mathbf{k})\rangle=\langle\eta_{yz}(-\mathbf{k})\eta_{yz}(\mathbf{k})\rangle=K(k)/V$.
With this assumption
\begin{equation}
\langle
v_s^2\rangle=\left(\frac{\hbar}{2m}\right)^2\int\frac{K(k)d^3k}{(2\pi)^3\xi_s^4}\frac{k_z^2(k^2-k_z^2)}{k^2(2k^2+k_z^2)^2}
\end{equation}
After integration over directions of $\mathbf{k}$ we arrive at:
\begin{equation}
\langle
v_s^2\rangle=\left(\frac{\hbar}{2m}\right)^2\int\frac{K(k)dk}{(2\pi)^2\xi_s^4}I,
\end{equation}
where
$I=2\int_0^1du\frac{u^2(1-u^2)}{(2+u^2)^2}=\frac{7}{\sqrt{2}}\arccos(\sqrt{\frac{2}{3}})-3\approx0.05$.
If the assumption $K(k)=const$ is used the integral over $k$
diverges linearly on the upper limit. For a crude estimation of the
diverging integral we can cut it at $k\sim(1/\xi_c)$ where the
assumption $K(k)=const$ breaks down, then $\langle v_s^2\rangle\sim
v_F^2(a/\xi_s)^2(1-P)$, here $a$ is interatomic distance. A more
refined treatment is based on the fact that at distances of the
order of $\xi_c$ aerogel has fractal dimensionality
\cite{parp2,halp2} and at large $k$ $K(k)\sim(1/k^D)$ with $D\simeq
1.7-1.9$, then the integral converges \cite{fom1}, but the result
depends on additional model assumptions or additional experimental
data  about the structure of aerogel.

\section{}

Integrals from equations (\ref{cor_l_isotropic_1}),
(\ref{cor_theta_parallel}) can be evaluated using cylindrical
coordinates ($k_z,~k_{\perp},~\varphi$). Integration on $k_z$  and
then on $\varphi$ yields zero-order Bessel function. The final
expressions are the following:
\begin{eqnarray}
\frac{\langle(\mathbf{l}(\mathbf{r}_2)-\mathbf{l}(\mathbf{r}_1))^2\rangle}{l_R^2}=\frac{1}{\xi_{LIM}}\left\{\frac{17}{2}\left[z^2+
\rho^2\right]^{\frac{1}{2}}+ \right.\nonumber
\\
\left.\frac{1}{6}\left[z^2+3\rho^2\right]^{\frac{1}{2}}-\frac{10\sqrt{6}}{3}\left[\frac{2}{3}z^2+\rho^2\right]^{\frac{1}{2}}-\right.\nonumber
\\
\left.16z\int\limits_0^{\infty}\frac{dk_{\perp}}{k_{\perp}}J_0({k_{\perp}\rho})\sinh\left(\frac{1-\sqrt{\frac{2}{3}}}{2}k_{\perp}z\right)
e^{-\frac{1+\sqrt{\frac{2}{3}}}{2}k_{\perp}z}\right\}\nonumber
\\
\end{eqnarray}
\begin{eqnarray}
\langle(\theta_{\|}(\mathbf{r}_2)-\theta_{\|}(\mathbf{r}_1))^2\rangle=\frac{1}{\xi_{LIM}}\left(\frac{9\sqrt{6}}{4}\left[\frac{2}{3}z^2+\rho^2\right]^{\frac{1}{2}}-\right.\nonumber
\\
\frac{11}{2}\left[z^2+\rho^2\right]^{\frac{1}{2}}+\nonumber
\\
\left.10z\int\limits_0^{\infty}\frac{dk_{\perp}}{k_{\perp}}J_0({k_{\perp}\rho})\sinh\left(\frac{1-\sqrt{\frac{2}{3}}}{2}k_{\perp}z\right)
e^{-\frac{1+\sqrt{\frac{2}{3}}}{2}k_{\perp}z}\right)\nonumber
\\
\end{eqnarray}
Evaluation of the integrals including Bessel functions for simple
cases $\mathbf{r}\|\mathbf{z}$ and $\mathbf{r}\perp \mathbf{z}$
integrals with the Bessel function are evaluated and the final
answers are given in the text above (Eqs.
(\ref{asymp_l_par})-(\ref{asymp_theta_per})).

\section{}
Here we present expressions for correlators in a form of multiple
integrals for the case of uniform stretching. The following
shorthand notations are used: $\sin\alpha=s_{\alpha}$,
$\cos\alpha=c_{\alpha}$,
\begin{eqnarray}
r_w(\theta,\varphi,\mathbf{r},p)=(xs_{\theta}c_{\varphi}+ys_{\theta}s_{\varphi}+zc_{\theta})p,\nonumber\\
\varepsilon(\theta,\varphi)=\frac{(\frac{1}{2}c^2_{\varphi}s_{2\theta}^2+c^2_{\theta}+2)^{1/2}}{(2c_{\theta}^4+5c_{\theta}^2+2)^{1/2}}.\nonumber
\end{eqnarray}
\begin{eqnarray}
\frac{\langle(\mathbf{l}(\mathbf{r})-\mathbf{l}(0))^2\rangle}{l_R^2}=\frac{1}{p\xi_{LIM}}\int
\frac{d\Omega}{4\pi} A(\theta, \varphi, \mathbf{r}),
\end{eqnarray}
where
\begin{eqnarray}
A(\theta,
\varphi,\mathbf{r})=\frac{1}{2}\left[\frac{1}{\varepsilon}\left(1-(1-
r_w\varepsilon)e^{-r_w\varepsilon}\right)G_1+\right.\nonumber
\\
\frac{1}{\varepsilon^3}\left(1-(1+
r_w\varepsilon)e^{-r_w\varepsilon}\right)G_2-\nonumber
\\
\left.\frac{1}{\varepsilon^5}\left(3-2r_w\varepsilon-(3+
r_w\varepsilon)e^{-r_w\varepsilon}\right)G_3\right],\nonumber
\\
G_1(\theta,\varphi)=\frac{2(3-s_{\theta}^4)^2+s_{\theta}^4c_{\theta}^4(1+c^2_{2\varphi})+\frac{1}{4}s_{2\varphi}^2s_{\theta}^2c_{\theta}^2}
{(2c_{\theta}^4+5c_{\theta}^2+2)^2},\nonumber\\
G_2(\theta,\varphi)=\frac{2(1+s_{\theta}^2)(3-s_{\theta}^4+c_{2\varphi}s_{\theta}^2c_{\theta}^2)}
{(2c_{\theta}^4+5c_{\theta}^2+2)^2},\nonumber\\
G_3(\theta)=\frac{(1+s_{\theta}^2)^2}{(2c_{\theta}^4+5c_{\theta}^2+2)^2}.\nonumber
\end{eqnarray}

\begin{eqnarray}
{\langle(\theta_{\|}(\mathbf{r})-\theta_{\|}(0))^2\rangle}=\frac{1}{p\xi_{LIM}}\int
\frac{d\Omega}{4\pi} B(\theta, \varphi, \mathbf{r})
\end{eqnarray}
\begin{eqnarray}
B(\theta,
\varphi,\mathbf{r})=\frac{1}{2}\left[\frac{1}{\varepsilon}\left(1-(1-
r_w\varepsilon)e^{-r_w\varepsilon}\right)H_1+\right.\nonumber
\\
\frac{1}{\varepsilon^3}\left(1-(1+
r_w\varepsilon)e^{-r_w\varepsilon}\right)H_2-\nonumber
\\
\left.\frac{1}{\varepsilon^5}\left(3-2r_w\varepsilon-(3+
r_w\varepsilon)e^{-r_w\varepsilon}\right)H_3\right],\nonumber
\\
H_1(\theta)=\frac{s_{\theta}^2c_{\theta}^2(c_{2\theta}+2)}
{(2c_{\theta}^4+5c_{\theta}^2+2)^2},\nonumber\\
H_2(\theta,\varphi)=\frac{2s_{\theta}^2c_{\theta}^2(c_{2\theta}+2)s_{\varphi}^2}
{(2c_{\theta}^4+5c_{\theta}^2+2)^2},\nonumber\\
H_3(\theta,\varphi)=\frac{s_{\theta}^2c_{\theta}^2s_{\varphi}^2}{(2c_{\theta}^4+5c_{\theta}^2+2)^2}.\nonumber
\end{eqnarray}
\begin{eqnarray}
{\langle(\theta_{u}(\mathbf{r})-\theta_{u}(0))^2\rangle}=\frac{1}{p\xi_{LIM}}\int
\frac{d\Omega}{4\pi} C(\theta, \varphi, \mathbf{r})\\
C(\theta,
\varphi,\mathbf{r})=\frac{1}{2}\left[\frac{1}{\varepsilon}\left(1-(1-
r_w\varepsilon)e^{-r_w\varepsilon}\right)J\right],\nonumber\\
J(\theta,\varphi)=\frac{(3-s_{\theta}^4+c_{\theta}^2s_{\theta}^2c_{2\varphi})^2+c_{\theta}^4s_{\theta}^4s_{2\varphi}^2}
{(2c_{\theta}^4+5c_{\theta}^2+2)^2}\nonumber
\end{eqnarray}

\bibliography{liter2017}

\providecommand{\noopsort}[1]{}\providecommand{\singleletter}[1]{#1}%
\begin{thebibliography}{25}%
\makeatletter
\providecommand \@ifxundefined [1]{%
 \@ifx{#1\undefined}
}%
\providecommand \@ifnum [1]{%
 \ifnum #1\expandafter \@firstoftwo
 \else \expandafter \@secondoftwo
 \fi
}%
\providecommand \@ifx [1]{%
 \ifx #1\expandafter \@firstoftwo
 \else \expandafter \@secondoftwo
 \fi
}%
\providecommand \natexlab [1]{#1}%
\providecommand \enquote  [1]{``#1''}%
\providecommand \bibnamefont  [1]{#1}%
\providecommand \bibfnamefont [1]{#1}%
\providecommand \citenamefont [1]{#1}%
\providecommand \href@noop [0]{\@secondoftwo}%
\providecommand \href [0]{\begingroup \@sanitize@url \@href}%
\providecommand \@href[1]{\@@startlink{#1}\@@href}%
\providecommand \@@href[1]{\endgroup#1\@@endlink}%
\providecommand \@sanitize@url [0]{\catcode `\\12\catcode `\$12\catcode
  `\&12\catcode `\#12\catcode `\^12\catcode `\_12\catcode `\%12\relax}%
\providecommand \@@startlink[1]{}%
\providecommand \@@endlink[0]{}%
\providecommand \url  [0]{\begingroup\@sanitize@url \@url }%
\providecommand \@url [1]{\endgroup\@href {#1}{\urlprefix }}%
\providecommand \urlprefix  [0]{URL }%
\providecommand \Eprint [0]{\href }%
\providecommand \doibase [0]{http://dx.doi.org/}%
\providecommand \selectlanguage [0]{\@gobble}%
\providecommand \bibinfo  [0]{\@secondoftwo}%
\providecommand \bibfield  [0]{\@secondoftwo}%
\providecommand \translation [1]{[#1]}%
\providecommand \BibitemOpen [0]{}%
\providecommand \bibitemStop [0]{}%
\providecommand \bibitemNoStop [0]{.\EOS\space}%
\providecommand \EOS [0]{\spacefactor3000\relax}%
\providecommand \BibitemShut  [1]{\csname bibitem#1\endcsname}%
\let\auto@bib@innerbib\@empty
%</preamble>
\bibitem [{\citenamefont {Larkin}(1970)}]{Lark}%
  \BibitemOpen
  \bibfield  {author} {\bibinfo {author} {\bibfnamefont {A.~I.}\ \bibnamefont
  {Larkin}},\ }\href@noop {} {\bibfield  {journal} {\bibinfo  {journal} {Zh.
  Eksp. Teor. Fiz.}\ }\textbf {\bibinfo {volume} {58}},\ \bibinfo {pages}
  {1466} (\bibinfo {year} {1970})},\ \translation{Sov. Phys. JETP \textbf{31},
  784 (1970)}\BibitemShut {NoStop}%
\bibitem [{\citenamefont {Imry}\ and\ \citenamefont {Ma}(1975)}]{IMRY_Ma}%
  \BibitemOpen
  \bibfield  {author} {\bibinfo {author} {\bibfnamefont {Y.}~\bibnamefont
  {Imry}}\ and\ \bibinfo {author} {\bibfnamefont {S.}~\bibnamefont {Ma}},\
  }\href@noop {} {\bibfield  {journal} {\bibinfo  {journal} {Phys. Rev. Lett.}\
  }\textbf {\bibinfo {volume} {35}},\ \bibinfo {pages} {1399} (\bibinfo {year}
  {1975})}\BibitemShut {NoStop}%
\bibitem [{\citenamefont {Chudnovsky}\ \emph {et~al.}(1986)\citenamefont
  {Chudnovsky}, \citenamefont {Saslow},\ and\ \citenamefont
  {Serota}}]{Chud1986}%
  \BibitemOpen
  \bibfield  {author} {\bibinfo {author} {\bibfnamefont {E.~M.}\ \bibnamefont
  {Chudnovsky}}, \bibinfo {author} {\bibfnamefont {W.~M.}\ \bibnamefont
  {Saslow}}, \ and\ \bibinfo {author} {\bibfnamefont {R.~A.}\ \bibnamefont
  {Serota}},\ }\href@noop {} {\bibfield  {journal} {\bibinfo  {journal} {Phys.
  Rev. B}\ }\textbf {\bibinfo {volume} {33}},\ \bibinfo {pages} {251} (\bibinfo
  {year} {1986})}\BibitemShut {NoStop}%
\bibitem [{\citenamefont {Imry}(1984)}]{IMRY_1984}%
  \BibitemOpen
  \bibfield  {author} {\bibinfo {author} {\bibfnamefont {Y.}~\bibnamefont
  {Imry}},\ }\href@noop {} {\bibfield  {journal} {\bibinfo  {journal} {J. Stat.
  Phys.}\ }\textbf {\bibinfo {volume} {34}},\ \bibinfo {pages} {849} (\bibinfo
  {year} {1984})}\BibitemShut {NoStop}%
\bibitem [{\citenamefont {Garanin}\ and\ \citenamefont
  {Chudnovsky}(2015)}]{Chud2015_1}%
  \BibitemOpen
  \bibfield  {author} {\bibinfo {author} {\bibfnamefont {D.}~\bibnamefont
  {Garanin}}\ and\ \bibinfo {author} {\bibfnamefont {E.~M.}\ \bibnamefont
  {Chudnovsky}},\ }\href@noop {} {\bibfield  {journal} {\bibinfo  {journal}
  {Eur. Phys. J. B}\ }\textbf {\bibinfo {volume} {88}},\ \bibinfo {pages} {81}
  (\bibinfo {year} {2015})}\BibitemShut {NoStop}%
\bibitem [{\citenamefont {Proctor}\ and\ \citenamefont
  {Chudnovsky}(2015)}]{Chud2015_2}%
  \BibitemOpen
  \bibfield  {author} {\bibinfo {author} {\bibfnamefont {T.~C.}\ \bibnamefont
  {Proctor}}\ and\ \bibinfo {author} {\bibfnamefont {E.~M.}\ \bibnamefont
  {Chudnovsky}},\ }\href@noop {} {\bibfield  {journal} {\bibinfo  {journal}
  {Phys. Rev. B}\ }\textbf {\bibinfo {volume} {91}},\ \bibinfo {pages} {140201}
  (\bibinfo {year} {2015})}\BibitemShut {NoStop}%
\bibitem [{\citenamefont {Porto}\ and\ \citenamefont {Parpia}(1995)}]{parp1}%
  \BibitemOpen
  \bibfield  {author} {\bibinfo {author} {\bibfnamefont {J.~V.}\ \bibnamefont
  {Porto}}\ and\ \bibinfo {author} {\bibfnamefont {J.~M.}\ \bibnamefont
  {Parpia}},\ }\href@noop {} {\bibfield  {journal} {\bibinfo  {journal} {Phys.
  Rev. Lett.}\ }\textbf {\bibinfo {volume} {74}},\ \bibinfo {pages} {4667}
  (\bibinfo {year} {1995})}\BibitemShut {NoStop}%
\bibitem [{\citenamefont {V.V.Dmitriev}\ \emph {et~al.}(2010)\citenamefont
  {V.V.Dmitriev}, \citenamefont {D.A.Krasnikhin}, \citenamefont {N.Mulders},
  \citenamefont {A.A.Senin}, \citenamefont {G.E.Volovik},\ and\ \citenamefont
  {A.N.Yudin}}]{Dmitriev2010}%
  \BibitemOpen
  \bibfield  {author} {\bibinfo {author} {\bibnamefont {V.V.Dmitriev}},
  \bibinfo {author} {\bibnamefont {D.A.Krasnikhin}}, \bibinfo {author}
  {\bibnamefont {N.Mulders}}, \bibinfo {author} {\bibnamefont {A.A.Senin}},
  \bibinfo {author} {\bibnamefont {G.E.Volovik}}, \ and\ \bibinfo {author}
  {\bibnamefont {A.N.Yudin}},\ }\href@noop {} {\bibfield  {journal} {\bibinfo
  {journal} {Pis`ma Zh. Eksp. Teor. Fiz.}\ }\textbf {\bibinfo {volume} {91}},\
  \bibinfo {pages} {669} (\bibinfo {year} {2010})},\ \translation{JETP Lett.
  \textbf{91}, 599 (2010)}\BibitemShut {NoStop}%
\bibitem [{\citenamefont {G.E.Volovik}(1996)}]{Volovik1996}%
  \BibitemOpen
  \bibfield  {author} {\bibinfo {author} {\bibnamefont {G.E.Volovik}},\
  }\href@noop {} {\bibfield  {journal} {\bibinfo  {journal} {Pis`ma Zh. Eksp.
  Teor. Fiz.}\ }\textbf {\bibinfo {volume} {63}},\ \bibinfo {pages} {301}
  (\bibinfo {year} {1996})},\ \translation{JETP Lett. \textbf{63}, 301,
  (1996)}\BibitemShut {NoStop}%
\bibitem [{\citenamefont {Sprague}\ \emph {et~al.}(1995)\citenamefont
  {Sprague}, \citenamefont {Haard}, \citenamefont {Kycia}, \citenamefont
  {Rand}, \citenamefont {Lee}, \citenamefont {Hamot},\ and\ \citenamefont
  {Halperin}}]{halp1}%
  \BibitemOpen
  \bibfield  {author} {\bibinfo {author} {\bibfnamefont {D.~T.}\ \bibnamefont
  {Sprague}}, \bibinfo {author} {\bibfnamefont {T.~M.}\ \bibnamefont {Haard}},
  \bibinfo {author} {\bibfnamefont {J.~B.}\ \bibnamefont {Kycia}}, \bibinfo
  {author} {\bibfnamefont {M.~R.}\ \bibnamefont {Rand}}, \bibinfo {author}
  {\bibfnamefont {Y.}~\bibnamefont {Lee}}, \bibinfo {author} {\bibfnamefont
  {P.~J.}\ \bibnamefont {Hamot}}, \ and\ \bibinfo {author} {\bibfnamefont
  {W.~P.}\ \bibnamefont {Halperin}},\ }\href@noop {} {\bibfield  {journal}
  {\bibinfo  {journal} {Phys. Rev. Lett.}\ }\textbf {\bibinfo {volume} {75}},\
  \bibinfo {pages} {661} (\bibinfo {year} {1995})}\BibitemShut {NoStop}%
\bibitem [{\citenamefont {Volovik}(2008)}]{vol2008}%
  \BibitemOpen
  \bibfield  {author} {\bibinfo {author} {\bibfnamefont {G.~E.}\ \bibnamefont
  {Volovik}},\ }\href@noop {} {\bibfield  {journal} {\bibinfo  {journal} {J.
  Low Temp. Phys.}\ }\textbf {\bibinfo {volume} {150}},\ \bibinfo {pages} {453}
  (\bibinfo {year} {2008})}\BibitemShut {NoStop}%
\bibitem [{\citenamefont {Fomin}(2016)}]{fom2016}%
  \BibitemOpen
  \bibfield  {author} {\bibinfo {author} {\bibfnamefont {I.~A.}\ \bibnamefont
  {Fomin}},\ }\href@noop {} {\bibfield  {journal} {\bibinfo  {journal} {Pis'ma
  Zh. Eksp. Teor. Fiz.}\ }\textbf {\bibinfo {volume} {104}},\ \bibinfo {pages}
  {18} (\bibinfo {year} {2016})},\ \translation{JETP Lett. \textbf{104}, 20
  (2016)}\BibitemShut {NoStop}%
\bibitem [{\citenamefont {Fomin}(2005)}]{fom_j}%
  \BibitemOpen
  \bibfield  {author} {\bibinfo {author} {\bibfnamefont {I.~A.}\ \bibnamefont
  {Fomin}},\ }\href@noop {} {\bibfield  {journal} {\bibinfo  {journal} {Journ.
  Phys. and Chem. of Solids}\ }\textbf {\bibinfo {volume} {66}},\ \bibinfo
  {pages} {1321} (\bibinfo {year} {2005})}\BibitemShut {NoStop}%
\bibitem [{\citenamefont {Vollhardt}\ and\ \citenamefont {Woelfle}(1990)}]{VW}%
  \BibitemOpen
  \bibfield  {author} {\bibinfo {author} {\bibfnamefont {D.}~\bibnamefont
  {Vollhardt}}\ and\ \bibinfo {author} {\bibfnamefont {P.}~\bibnamefont
  {Woelfle}},\ }\href@noop {} {\emph {\bibinfo {title} {The Superfluid Phases
  of Helium 3}}}\ (\bibinfo  {publisher} {Tailor and Francis, London, New York,
  Phyladelphia},\ \bibinfo {year} {1990})\BibitemShut {NoStop}%
\bibitem [{\citenamefont {Larkin}\ and\ \citenamefont
  {Ovchinnikov}(1971)}]{LO}%
  \BibitemOpen
  \bibfield  {author} {\bibinfo {author} {\bibfnamefont {A.~I.}\ \bibnamefont
  {Larkin}}\ and\ \bibinfo {author} {\bibfnamefont {Y.~N.}\ \bibnamefont
  {Ovchinnikov}},\ }\href@noop {} {\bibfield  {journal} {\bibinfo  {journal}
  {Zh. Eksp. Teor. Fiz.}\ }\textbf {\bibinfo {volume} {61}},\ \bibinfo {pages}
  {1221} (\bibinfo {year} {1971})},\ \translation{Sov. Phys. JETP \textbf{34},
  651 (1971)}\BibitemShut {NoStop}%
\bibitem [{\citenamefont {Thuneberg}(1998)}]{thuneb_1998}%
  \BibitemOpen
  \bibfield  {author} {\bibinfo {author} {\bibfnamefont {E.~V.}\ \bibnamefont
  {Thuneberg}},\ }\href@noop {} {\bibfield  {journal} {\bibinfo  {journal}
  {Quasiclassical methods in superconductivity and superfluidity, Verditz 96}\
  ,\ \bibinfo {pages} {53}} (\bibinfo {year} {1998})}\BibitemShut {NoStop}%
\bibitem [{\citenamefont {Surovtsev}\ and\ \citenamefont
  {Fomin}(2008)}]{SF2008}%
  \BibitemOpen
  \bibfield  {author} {\bibinfo {author} {\bibfnamefont {E.~V.}\ \bibnamefont
  {Surovtsev}}\ and\ \bibinfo {author} {\bibfnamefont {I.~A.}\ \bibnamefont
  {Fomin}},\ }\href@noop {} {\bibfield  {journal} {\bibinfo  {journal} {J. Low
  Temp. Phys.}\ }\textbf {\bibinfo {volume} {150}},\ \bibinfo {pages} {487}
  (\bibinfo {year} {2008})}\BibitemShut {NoStop}%
\bibitem [{\citenamefont {Rainer}\ and\ \citenamefont {Vuorio}(1977)}]{Rainer}%
  \BibitemOpen
  \bibfield  {author} {\bibinfo {author} {\bibfnamefont {D.}~\bibnamefont
  {Rainer}}\ and\ \bibinfo {author} {\bibfnamefont {M.}~\bibnamefont
  {Vuorio}},\ }\href@noop {} {\bibfield  {journal} {\bibinfo  {journal} {J.
  Phys. C: Solid State Phys.}\ }\textbf {\bibinfo {volume} {10}},\ \bibinfo
  {pages} {3093} (\bibinfo {year} {1977})}\BibitemShut {NoStop}%
\bibitem [{\citenamefont {Kunimatsu}\ \emph {et~al.}(2007)\citenamefont
  {Kunimatsu}, \citenamefont {Sato}, \citenamefont {Izumina}, \citenamefont
  {Matsubara}, \citenamefont {Sasaki}, \citenamefont {Kubota}, \citenamefont
  {Ishikawa}, \citenamefont {Mizusaki},\ and\ \citenamefont {Bunkov}}]{knmts}%
  \BibitemOpen
  \bibfield  {author} {\bibinfo {author} {\bibfnamefont {T.}~\bibnamefont
  {Kunimatsu}}, \bibinfo {author} {\bibfnamefont {T.}~\bibnamefont {Sato}},
  \bibinfo {author} {\bibfnamefont {K.}~\bibnamefont {Izumina}}, \bibinfo
  {author} {\bibfnamefont {A.}~\bibnamefont {Matsubara}}, \bibinfo {author}
  {\bibfnamefont {Y.}~\bibnamefont {Sasaki}}, \bibinfo {author} {\bibfnamefont
  {M.}~\bibnamefont {Kubota}}, \bibinfo {author} {\bibfnamefont
  {O.}~\bibnamefont {Ishikawa}}, \bibinfo {author} {\bibfnamefont
  {T.}~\bibnamefont {Mizusaki}}, \ and\ \bibinfo {author} {\bibfnamefont
  {Y.~M.}\ \bibnamefont {Bunkov}},\ }\href@noop {} {\bibfield  {journal}
  {\bibinfo  {journal} {Pis'ma Zh. Eksp. Teor. Fiz.}\ }\textbf {\bibinfo
  {volume} {86}},\ \bibinfo {pages} {244} (\bibinfo {year} {2007})},\
  \translation{JETP Lett. \textbf{86}, 216 (2007)}\BibitemShut {NoStop}%
\bibitem [{\citenamefont {Fomin}\ and\ \citenamefont
  {Surovtsev}(2015)}]{FS2015}%
  \BibitemOpen
  \bibfield  {author} {\bibinfo {author} {\bibfnamefont {I.~A.}\ \bibnamefont
  {Fomin}}\ and\ \bibinfo {author} {\bibfnamefont {E.~V.}\ \bibnamefont
  {Surovtsev}},\ }\href@noop {} {\bibfield  {journal} {\bibinfo  {journal}
  {Phys. Rev. B}\ }\textbf {\bibinfo {volume} {92}},\ \bibinfo {pages} {064509}
  (\bibinfo {year} {2015})}\BibitemShut {NoStop}%
\bibitem [{\citenamefont {Elbs}\ \emph {et~al.}(2008)\citenamefont {Elbs},
  \citenamefont {Bunkov}, \citenamefont {Collin}, \citenamefont {Godfrin},\
  and\ \citenamefont {Volovik}}]{Elbs}%
  \BibitemOpen
  \bibfield  {author} {\bibinfo {author} {\bibfnamefont {J.}~\bibnamefont
  {Elbs}}, \bibinfo {author} {\bibfnamefont {Y.~M.}\ \bibnamefont {Bunkov}},
  \bibinfo {author} {\bibfnamefont {E.}~\bibnamefont {Collin}}, \bibinfo
  {author} {\bibfnamefont {H.}~\bibnamefont {Godfrin}}, \ and\ \bibinfo
  {author} {\bibfnamefont {G.~E.}\ \bibnamefont {Volovik}},\ }\href@noop {}
  {\bibfield  {journal} {\bibinfo  {journal} {Phys. Rev. Lett.}\ }\textbf
  {\bibinfo {volume} {100}},\ \bibinfo {pages} {215304} (\bibinfo {year}
  {2008})}\BibitemShut {NoStop}%
\bibitem [{\citenamefont {Li}\ \emph {et~al.}(2013)\citenamefont {Li},
  \citenamefont {Zimmerman}, \citenamefont {Pollanen}, \citenamefont {Collett},
  \citenamefont {Gannon},\ and\ \citenamefont {Halperin}}]{Halperin_orient}%
  \BibitemOpen
  \bibfield  {author} {\bibinfo {author} {\bibfnamefont {J.~I.~A.}\
  \bibnamefont {Li}}, \bibinfo {author} {\bibfnamefont {A.~M.}\ \bibnamefont
  {Zimmerman}}, \bibinfo {author} {\bibfnamefont {J.}~\bibnamefont {Pollanen}},
  \bibinfo {author} {\bibfnamefont {C.~A.}\ \bibnamefont {Collett}}, \bibinfo
  {author} {\bibfnamefont {W.~J.}\ \bibnamefont {Gannon}}, \ and\ \bibinfo
  {author} {\bibfnamefont {W.~P.}\ \bibnamefont {Halperin}},\ }\href@noop {}
  {\bibfield  {journal} {\bibinfo  {journal} {J. Low Temp. Phys.}\ }\textbf
  {\bibinfo {volume} {175}},\ \bibinfo {pages} {31} (\bibinfo {year}
  {2013})}\BibitemShut {NoStop}%
\bibitem [{\citenamefont {Porto}\ and\ \citenamefont {Parpia}(1999)}]{parp2}%
  \BibitemOpen
  \bibfield  {author} {\bibinfo {author} {\bibfnamefont {J.~V.}\ \bibnamefont
  {Porto}}\ and\ \bibinfo {author} {\bibfnamefont {J.~M.}\ \bibnamefont
  {Parpia}},\ }\href@noop {} {\bibfield  {journal} {\bibinfo  {journal} {Phys.
  Rev. B}\ }\textbf {\bibinfo {volume} {59}},\ \bibinfo {pages} {14583}
  (\bibinfo {year} {1999})}\BibitemShut {NoStop}%
\bibitem [{\citenamefont {Halperin}\ \emph {et~al.}(2008)\citenamefont
  {Halperin}, \citenamefont {Choi}, \citenamefont {Davis},\ and\ \citenamefont
  {Polanen}}]{halp2}%
  \BibitemOpen
  \bibfield  {author} {\bibinfo {author} {\bibfnamefont {W.~P.}\ \bibnamefont
  {Halperin}}, \bibinfo {author} {\bibfnamefont {H.}~\bibnamefont {Choi}},
  \bibinfo {author} {\bibfnamefont {J.~P.}\ \bibnamefont {Davis}}, \ and\
  \bibinfo {author} {\bibfnamefont {J.}~\bibnamefont {Polanen}},\ }\href@noop
  {} {\bibfield  {journal} {\bibinfo  {journal} {J. Phys. Soc. Jpn.}\ }\textbf
  {\bibinfo {volume} {77}},\ \bibinfo {pages} {111002} (\bibinfo {year}
  {2008})}\BibitemShut {NoStop}%
\bibitem [{\citenamefont {Fomin}(2008)}]{fom1}%
  \BibitemOpen
  \bibfield  {author} {\bibinfo {author} {\bibfnamefont {I.~A.}\ \bibnamefont
  {Fomin}},\ }\href@noop {} {\bibfield  {journal} {\bibinfo  {journal} {Pis'ma
  Zh. Eksp. Teor. Fiz.}\ }\textbf {\bibinfo {volume} {88}},\ \bibinfo {pages}
  {65} (\bibinfo {year} {2008})},\ \translation{JETP Lett. \textbf{88}, 59
  (2008)}\BibitemShut {NoStop}%
\end{thebibliography}%

\end{document}